\documentstyle[preprint,eqsecnum,aps,graphicx]{revtex}
\tightenlines
\begin{document}
\draft
\title{Velocity Field Distributions Due to Ideal Line Vortices}
\author{Thomas S.\ Levi and David C.\ Montgomery}
\address{Department of Physics and Astronomy\\
Dartmouth College, Hanover, NH\quad 03755-3528}
\date{\today}
\maketitle
\begin{abstract}
We evaluate numerically the velocity field distributions produced by a bounded, two-dimensional fluid model consisting of a collection of parallel ideal line vortices. We sample at many spatial points inside a rigid circular boundary. We focus on ``nearest neighbor'' contributions that result from vortices that fall (randomly) very close to the spatial points where the velocity is being sampled. We confirm that these events lead to a non-Gaussian high-velocity ``tail'' on an otherwise Gaussian distribution function for the Eulerian velocity field. We also investigate the behavior of distributions that do not have equilibrium mean-field probability distributions that are uniform inside the circle, but instead correspond to both higher and lower mean-field energies than those associated with the uniform vorticity distribution. We find substantial differences between these and the uniform case.

\end{abstract}
\pacs{}


\section{Introduction}
\label{sec-intro}
Study of the hydrodynamics of ideal line vortices goes back at least as far as Helmholtz in the 19th century, and was developed in the 20th by Lin \cite{lin} and Onsager \cite{onsager}, who first made the dynamical system an object of statistical mechanical inquiry. The system appeared in plasma physics when Taylor and McNamara \cite{taylor,mont72} calculated the Bohm-like coefficients of self-diffusion for a strongly-magnetized, two-dimensional, electrostatic guiding-center plasma model, a system whose mathematical description becomes identical with that of the ideal line vortex system under appropriate substitutions; the fact that these diffusion coefficients were inversely proportional to the first power of the magnetic field, even in thermal equilibrium, was startling.

The system is one for which interesting statistical-mechanical and fluid-mechanical questions can be asked, but must be asked with care, for two reasons. First, viscous effects have never been fully included in the model, although some forms of Navier-Stokesian behavior have on occasion been observed for it. Secondly, no classical, extensive ``thermodynamic limit'' exists for the system in the conventional sense, and the partition function, even for the case in which there is no overall net vorticity, does not in general exist in the infinite volume limit \cite{kiessling}. None of the standard machinery of equilibrium statistical mechanics can be trusted completely without re-examination.

One question that can be asked, motivated in part by various probability distribution function measurements for turbulent fluid velocities that have been made in recent years, concerns the distribution of the velocity field at a fixed point in space, one at which no vortex necessarily resides. The field in question is one that is produced by all the vortices. This is a close analogue of the question of the probability distribution of the vector gravitational field due to a large collection of point masses, a question addressed in detail by Chandrasekhar in 1943 \cite{chandra}. Under the assumption that the point masses in three dimensions are uniformly distributed and uncorrelated, the resulting Holtsmark distribution has many non-standard properties, including the divergence of some of its low-order moments: a consequence of the long range of the inverse-square force field and the fact that point masses (or charges) each have an infinite ``self energy'' that reflects itself in the total force field when the single-particle contributions are combined additively.

In a recent interesting paper \cite{kuvshinov}, Kuvshinov and Schep considered the statistics of the velocity field of a large but finite number of ideal line vortices inside a circular boundary (see also the paper of Chukbar \cite{chukbar}, which is of some importance). They assumed uniformly distributed and uncorrelated line vortices of a single sign of vorticity. They noted that the Holtsmark-style treatment carried out by Chandrasekhar for the three-dimensional case contained a divergent integral in two dimensions, and so was not immediately applicable. They then performed repeated numerical measurements of the two-dimensional (2D) velocity field, near the center of the circular boundary, that resulted from uncorrelated random distributions of large numbers of vortices, thrown at each trial into the circular boundary without correlation and without any mean density variation.

The most interesting result of Kuvshinov and Schep was thus an ``experimentally'' determined probability distribution for the velocity which seemed to split naturally into two parts: a Gaussian distribution for the lower velocities and high-energy ``tails'' for the larger velocities that fell off approximately as the third power of the fluctuating velocity. (Here, ``fluctuating'' velocities are interpreted to mean those with the mean-field rigid rotation associated with the uniform vorticity density distribution subtracted out.)  They hypothesized that the approximate inverse third-power dependence of the tail was a consequence of occasional ``near neighbor'' contributions, in which one vortex found itself very close to the point where the velocity field was being sampled, and generalized a three-dimensional ``nearest-neighbor'' algebraic argument of Chandrasekhar's \cite{chandra} to account for this high-velocity power law contribution. In a rather different continuum model, something not totally dissimilar had previously been reported by Jimenez \cite{jimenez}.

We have in this paper repeated certain features of Kuvshinov and Schep's numerical experiment, and have attempted to modify and amplify it in a variety of ways. (1) We have inserted an ideal, perfectly-reflecting wall boundary at the radius of the confining circle by changing the Green's function to one that, by the method of images, guarantees the vanishing of all radial velocities at the boundary \cite{lundgren}, rather than using the inverse logarithmic Green's function appropriate to the unbounded region. (2) We have, upon finding the non-Gaussian high-velocity tails in the probability distribution function, implemented a program that searches numerically for near neighbor contributions to the locally measured velocity field, and when it finds one, deletes its contribution to the local velocity field. We find that as a consequence, the high-velocity tails disappear, thus reinforcing the conjecture of Ref.\ \cite{kuvshinov}. (3) We study the velocity field away from the origin, to determine how representative of the entire spatial volume the velocity field sampled at the center is. (4) Finally, we allow the mean vorticity density with which the vortex particles are distributed to vary, and rather than placing them randomly with a spatially-uniform mean-field distribution, we weight their locations with a probability distribution function that depends exponentially upon a mean-field stream function and has a temperature that can be positive or negative \cite{lundgren,montjoyce73}. The equilibrium statistical mechanics of the ideal line vortex system has undergone considerable development since it was introduced (e.g., \cite{lundgren,montjoyce73,montjoyce74,ting}, and references therein) and we take advantage of results which we will not go into full detail describing here. We note only that the pairwise, additive Coulomb potentials, summed over all the pairs in the system, are an ideal invariant dynamically which can take on virtually any value and which determines the single-time thermal-equilibrium probability distributions of all particles. Only one value of this energy is represented by the uniform distribution. We find significant differences in the velocity field statistics that result from total mean energies that are significantly higher or lower than those associated with the unfiorm (rigidly-rotating) mean-field distribution.

In Sec.\ \ref{sec-nuts}, we describe the comptutational procedure and the results for the uniform mean-field vorticity density distribution for points near the center of the circle, with an emphasis on non-Gaussian, high-velocity ``tails'' that appear in the probability distribution function for the velocity. In Sec.\ \ref{neighbors}, we introduce a cutoff below which ``near neighbor'' contributions to the velocity field are locally removed, and derive an analytic expression for the contribution of very near neighbors to the local velocity field distribution. Sec.\ \ref{sec-resu} discusses the statistics of the velocity field for the uniform density distribution away from the center of the container and near the boundary. Sec.\ \ref{sec-prob} is devoted to the case in which the mean number density of vortices is not uniform, but rather follows from a self-consistent, mean-field theory which permits the study of high and low energy states, relative to the uniform density state. Sec.\ \ref{sec-resp} presents the results for the non-uniform mean-field distributions. Sec.\ \ref{sec-conclusions} summarizes the results and indicates possible future directions for further investigations.

\section{General Procedure}
\label{sec-nuts}
In a point vortex model, where each vortex has strength $\kappa _j$ the flow is two-dimensional in the $(x,y)$ plane, has only $x$ and $y$ components, and is given by
\begin{equation}
\label{v-field}
{\bf v(r)} = \sum _j \kappa _j \nabla \times (G({\bf r,r_j}){\bf e_z})
\end{equation}
Here ${\bf e_z}$ is the unit vector pointing perpendicular to the plane of the spatial variation of the fluid,
$G$ is the Green's function that relates the vorticity to the stream function,
and the sum is over all (two-dimensional) vortex positions ${\bf r_j}$. Thus, we see that the 
velocity at
a given point is due to all the vortices not at that point. For a 
two-dimensional
fluid, in a rigid, circular container of radius $R$, the boundary condition is
that the normal component of ${\bf v}$ go to zero at the wall. The 
appropriate Green's function to choose is \cite{lundgren}
\begin{equation}
\label{green-fcn}
G({\bf r,r'})= \frac{1}{2 \pi} \ln (|{\bf r-r'}|) - \frac{1}{2 \pi} \ln \Bigl
(\Big|{\bf r} - \frac{R^2}{r'^2} {\bf r'} \Big| \frac{r'}{R} \Bigr)
\end{equation}
Where here we have replaced ${\bf r_j}$ with ${\bf r'}$. Using Eq.\ (\ref{v-field}) we get
\begin{mathletters}
\begin{equation}
\label{vr}
v_r = \frac{\kappa}{2 \pi} \biggr( \frac{R^2 r' \sin \theta _{12} }{r^2 r'^2 + 
R^4 - 2 R^2 r r' \cos \theta _{12} } - \frac{r' \sin \theta _{12}}{r^2 +r'^2
 - 2rr' \cos \theta _{12} } \biggr)
\end{equation}
\begin{equation}
\label{vt}
v_\theta = \frac{\kappa}{2 \pi} \biggr(- \frac{r r'^2 - R^2 r' \cos \theta _{12} }
{r^2 r'^2 + R^4 - 2 R^2 r r' \cos \theta _{12} } + \frac{r-r' \cos \theta_ {12}}
{r^2 +r'^2 - 2rr' \cos \theta _{12} } \biggr)
\end{equation}
\end{mathletters}
Where $v_r$ and $v_\theta$ represent the $r$ and $\theta$ components of velocity
due to one point vortex of strength $\kappa$, and $\theta_{12}$ is the angle 
between the radii to the point where the velocity is measured and the position of the vortex. For each component the terms with $R$ represent the terms that are a result of the finite boundary.

All quantities will be expressed throughout in terms of dimensionless variables appropriate to the model. Since the Euler dynamics contain no viscosity, all quantities in the  dynamics before non-dimensionalization contain only combinations of lengths and times, or equivalently, velocities and times, so units are not of great significance. For a convenient basic unit of length, we may take the mean nearest-neighbor separation in a uniform vorticity benchmark case divided by $\pi ^{1/2}$ and for the basic unit of velocity, the speed with which an isolated vortex of strength $2 \pi$ will rotate the fluid in which it is imbedded at unit length distance from the vortex.

The general procedure we use is to place a large number, $N$, of vortices of
strength $\kappa = 2 \pi$ into a circular region of radius $R$ using a random number generator and study the statistics of the resulting velocity field. Specifically, we examine the probability distribution for the scalar fluctuating velocity ${\bf |w|} = w$, where ${\bf w} = {\bf v- V}$, and ${\bf V}$ is the mean-field velocity. Let $f(w)d{\bf w}$ be the probability that the velocity is in the area element (in velocity space) $d {\bf w}$ centered at ${\bf w}$. We are here assuming that the distribution is isotropic in velocity space, which is confirmed by our numerics everywhere except in a very thin layer near the radial boundary. We wish to switch to a one-dimensional integral, which is done by letting $F(w)dw= 2 \pi w f(w) dw$. The resulting
distribution $F(w)$ is normalized such that $\int _{o} ^{\infty} F(w) d w = 1$.
Our graphs contain a numerically obtained $F(w)$. The procedure for obtaining this $F$ is to first run a series of trials, each trial representing a set of random choices for the vortex positions inside the circle. For the uniform vorticity density case, we have run $3000$ trials. Then, we record a velocity value at each point sampled in the circle. Here we have sampled at 50 points separated by uniform intervals from $r=0$ to $r=399$, where $R=400$ and $N=1.6 \times 10^5$. We then bin the velocities using a histogram with uniform spacing between bins. This procedure gives us an unnormalized probability distribution for $f$. To get from this step to the actual $F$ plotted requires two steps: (1) We first multiply each bin value by the $w$ at the center of its bin. (2) We normalize the result using a trapezoidal numerical integration, so that, numerically $\int F(w)dw = 1$. It is easiest to see the probability distribution's behavior on a natural log plot, so we plot $\ln (F(w)/w)$ versus $w^2$. The error bars are one standard deviation of the mean in length above and below; namely, we calculate the standard deviation of $\ln (F/w) $ and then divide by the square root of the number of actual events that fall into that histogram bin. We present two graphs for each point sampled in the uniform case: (1) A graph that includes all numerical events. (2) A graph with the ``nearest-neighbor'' events subtracted out. The subtraction procedure is defined relatively simply and somewhat arbitrarily. At each point sampled, the program records the distance to all of the vortices placed in the region. If the distance $d$ is such that $d < 0.65$ then that event is deleted from the distribution for that point only. That is, if there is a nearest neighbor event recorded at $ r = 200$ for example, its removal will {\it not} affect the resulting distribution at any other point. The resulting distribution can be thought of as the probability distribution if there were never any ``nearest-neighbor'' events. In each plot, the solid line is a best-fit Gaussian given by \cite{kuvshinov}:
\begin{equation}
\label{anal-F}
F(w) = \frac {w}{\overline{w}^2} \exp (-\frac{w^2}{2 \overline{w}^2})
\end{equation}
Where $ \overline{w} $ is a measure of the average velocity and is numerically
determined for a best fit. The dashed line represents an analytical expression 
for near neighbor contributions in the bounded case which will be calculated below.  

\section{Nearest Neighbors}
\label{neighbors}
Here we follow the general procedure of Chandrasekhar \cite{chandra}, but carry
it out in two dimensions and for a general mean-field vorticity density $n(r)$
to get an analytical expression for nearest neighbor events. Let $F_n (r')dr'$ represent the probability that that the nearest neighbor lies between $r'$ and $r'+dr'$. This probability must be equal to the probability that no neighbors are interior to $r'$ times the probability that a particles does exist in the circular shell between $r'$ and $r'+dr'$. Thus $F_n (r')$ must satisfy \cite{chandra}
\begin{equation}
\label{int-F}
F_n (r') = \Bigl(1 - \int _{0} ^{r'} F_n (r) dr \Bigr) 2 \pi r' n(r')
\end{equation}
where $r'$ is the distance to the nearest neighbor. Differentiating both sides, 
we get a differential equation for $F_n$
\begin{equation}
\label{diff-F}
\frac{d}{dr'} \biggl( \frac{F_n (r')}{2 \pi r' n(r')}\biggr) = -2 \pi r' n(r')
\frac{F_n (r')}{2 \pi r' n(r')}
\end{equation}
This equation is not hard to solve; its solution is
\begin{equation}
\label{sol-F}
F_n (r') = 2 \pi r' n(r') C \exp \Bigl(- 2 \pi \int _{0} ^{r'} n(r) r dr \Bigr)
\end{equation}
Where C is a normalization constant such that $ \int _{0} ^{R} F_n (r')dr' = 1$. In
general, $ C \sim \frac{1}{1-e^{-N}} $, and since $N \gg 1$, $C \cong 1$. In
particular, for $n=constant$ and small $r'$, we get
\begin{equation}
\label{consn-F}
F_n (r')= 2 \pi r' n \exp(- \pi n r'^2) \cong 2 \pi r' n
\end{equation}
Using $ w = \frac {\kappa r'}{2 \pi} \Bigl( \frac{1}{r'^2} - \frac{1}{R^2}
\Bigr)$ which is exact at the origin ($r=0$), and a good approximation at points not at
the origin, we get
\begin{equation}
\label{F(w)}
F_n (w) = 2 \pi r' (w) n \frac{dr'}{dw}
\end{equation}
This $F_n (w)$ will be plotted as a dashed line when exhibiting the measured velocity distribution vs.\ $w$.

\section{Results for Uniform Vorticity Density Case}
\label{sec-resu}

Figs.\ \ref{fig1} and \ref{fig2} display results for the numerically determined velocity distribution for the uniform mean-field vorticity density runs, a total of $3000$ trials. Fig.\ \ref{fig1} shows results of sampling at $r=0$, and Fig.\ \ref{fig2} at $r=399$, quite close to the wall. At intermediate points, the results are quite similar to those at $r=0$. 

In Figs.\ \ref{fig1}a and \ref{fig1}b, the solid line represents the Gaussian, Eq.\ (\ref{anal-F}), with the same mean-square velocity fluctuation. The dashed line represents the nearest neighbor contribution, as predicted by Eq.\ (\ref{F(w)}). The ``experimentally'' determined points are shown with their associated error bars, estimated as described in Sec.\ \ref{sec-nuts}. Fig.\ \ref{fig1}a shows the results for the raw data, with no ``nearest neighbor'' events removed. Fig.\ \ref{fig1}b (the lower figure) shows the results of deleting the nearest neighbor events. The reason no data points appear above $w^2$ of about $85$ is that all the computed points above that value contain nearest neighbor events. A similar set of statements applies to Figs.\ \ref{fig2}a,b, which are for the radius $r=399$. In both cases, it appears that the high-velocity events are reasonably well predicted by Eq.\ (\ref{F(w)}). In both cases, the Gaussian (\ref{anal-F}) is clearly a good approximation only for the lower values of $w$.

Fig.\ \ref{fig3} shows the distribution of the numerically-obtained magnitude of the radial component of velocity as a function of $r$. The intent is to assess the effect of the rigid boundary at $r=400$, the location of the wall. It will be seen that the decrease of the radial velocities is significant only within a relatively thin boundary layer near the wall. If the vortex dynamics were allowed to evolve in time, it is expected that the boundary layer would persist, but might acquire dimensions not necessarily the same as observed for the purely random distribution.

Summarizing, we conclude that for the case in which the uniform mean-field vorticity density applies, there are indeed non-Gaussian tails present in the probability distributions, and we confirm the conjecture of Kuvshinov and Schep that they may be explained as the result of nearest-neighbor contributions. Only near the radial boundary does its presence result in any significant departure from the statistics observed in the interior, for this case.

\section{Non-Uniform Mean-Field Vorticities: ``Most Probable'' Distributions}
\label{sec-prob}
Up to this point, we have considered only the case of the uniform probability distribution for vortices. However, a much wider variety of thermal equilibrium states is possible for ideal line vortices, considered as a dynamical system (\cite{onsager,mont72,kiessling,lundgren,montjoyce73,montjoyce74,ting,matthaeus,montmatt,schep}, and references therein). The Hamiltonian or energy of the system is equivalent to the Coulomb energies of the pairs of interacting line vortices, summed over all the pairs, and is a constant of the motion for these boundary conditions.  More extensive investigations have been carried out for the two-species case than for the present one-species case, but one species may equally well be considered. The preceding results do not apply to any value of the energy expectation (which is determined by the initial conditions chosen when the system is considered dynamically) except the one associated with the completely uniform mean-field distribution. For either higher or lower energies, the thermal equilibrium, mean-field, one-body distribution is not spatially uniform. It is concentrated toward $r=0$ for higher energies, and around the rim for lower ones. In this Section, we provide an expression for the probability distribution for these higher and lower energy cases, referring to the rather extensive cited literature for the formalism and justification (\cite{lundgren,montjoyce73,montjoyce74,ting,matthaeus,montmatt,schep}, and references therein).

We find the mean fields from solving the one-species analogue of the ``$\sinh$-Poisson'' equation,
\begin{equation}
\label{liouville}
\nabla^2 \psi= - \omega = -e^{-\alpha-\beta\psi}
\end{equation}
where $\psi$ is the ``most probable'' stream function, and $\omega$ is its associated mean-field vorticity distribution. In the present case, it will be assumed that the relevant solutions are symmetric with respect to rotations about $r=0$.

Eq.\ ({\ref{liouville}) is to be solved subject to the constrainst that ${\cal E}=\frac{1}{2} \int ( \nabla \psi )^2 d^{2} x$ and $\Omega = - \int \nabla^2 \psi d^{2} x$, where ${\cal E}$ is the mean-field energy, and $\Omega$ is the total vorticity. If we assume $\psi$ is a function of radius only,  Eq.\ (\ref{liouville}) becomes simply $\frac{1}{r} \frac {d}{dr} r \frac{d \psi}{dr} = - \omega = - e^{-\alpha - \beta \psi}$, which is sometimes called Liouville's equation and has been widely studied (e.g.,\cite{montturn}).

We may solve the equation for $\psi$ by writing $\omega = c_1 / (1+c_{2} r^2)^2$. Taking the Laplacian of the natural logarithm, we get
\begin{equation}
\label{ln}
\frac{1}{r} \frac{d}{dr} r \frac{d \psi}{dr} = \frac{8c_{2} }{\beta (1+c_{2}
r^{2})^2}=- \omega = -\frac{c_1}{(1+c_{2} r^{2})^2}
\end{equation}
The equality demands that $c_1 = -8 c_2 / \beta$.  Inserting the expression into the constraint equations, we find that
\begin{mathletters}
\begin{equation}
\label{cons1_sol}
\Omega = - \frac{8 \pi}{\beta} \frac{c_{2} R^2}{1+c_{2} R^2}
\end{equation}
\begin{equation}
\label{cons2_sol}
{\cal E} = \frac{8 \pi}{\beta ^2} \biggl[\ln (1+c_{2} R^2) - \frac{c_{2} R^2}
{1+c_{2} R^2}\biggr]
\end{equation}
\end{mathletters}
The goal is to solve Eqs.\ (\ref{cons1_sol}) and (\ref{cons2_sol}) for the constants $c_2$ and $\beta$. The result is
\begin{equation}
\label{E/W^2}
\frac{{\cal E}}{\Omega ^2} = \frac{1}{8 \pi} \frac{(1+c_{2} R^2)^2}{(c_{2} 
R^2)^2}\biggl[\ln(1+c_{2} R^2) - \frac{c_{2} R^2}{1+c_{2} R^2} \biggr]
\end{equation}
which must be solved numerically for $c_2$ in terms of $\Omega$ and ${\cal E}$. The result is $\beta = - \frac{8 \pi}{\Omega} \frac{c_2 R^2}{1+ c_2 R^2}$ and $\omega = \frac{\Omega}{\pi R^2} \frac{1+c_2 R^2}{(1+c_2 r^2)^2}$, where $c_2$ is given by Eq.\ (\ref{E/W^2}). We have now expressed the mean-field vorticity directly in terms of energy and vorticity.
It follows that when placing vortices ``randomly'' into the circular region for numerical trials, we should weight their placements by a probability distribution that wil lead to the correct $\omega$ in the mean-field limit. That is,
\begin{equation}
\label{prob_fcn}
p(r, \theta) r dr = \frac{r}{\pi R^2} \frac{1+c_2 R^2}{(1+c_2 r^2)^2} dr
\end{equation}
Here, the radial probability density $p$ is normalized such that $\int p(r, \theta) r dr d \theta = 1$.  The spatially uniform case treated previously corresponds to the case $c_2 \rightarrow 0$, in which case we get ${\cal E}_0 = \Omega ^2 / 8 \pi $. The nearest neighbor formula must be modified to
\begin{equation}
\label{nonunif-nearest}
F_n (w) = 2 r'(w) \frac{N}{R^2} \frac{1+ c_2 R^2}{(1 + c_2 r'^2 (w) )^2} 
\frac{dr'}{dw} \exp \biggl( -(1+ c_2 R^2) \frac{N}{R^2} \frac{r'^2 (w)}
{1+ c_2 r'^2 (w) } \biggr)
\end{equation}

\section{Results for Non-Uniform trials}
\label{sec-resp}

As might be expected, noticeable differences occur when the mean-field vorticity is a function of radius.  First, the mean azimuthal velocity no longer corresponds to a rigid rotation, and the fluctuating velocity must be referred to it locally. Qualitatively, it might be expected that the higher energy trials will produce more nearest-neighbor events, at constant mean density over the whole circle, and hence a more intense velocity fluctuation spectrum, and the opposite for the lower energy cases.  That seems to be what happens.

We conducted two runs of $1790$ trials each, with $N = 1.6 \times 10^5$ and $R=400$, as before. One of the sets of trials corresponded to mean-field energy ${\cal E} = 4{\cal E}_0$ and the other set to ${\cal E} = {\cal E}_0 / 4$. Fig.\ \ref{fig4} shows the mean probability distribution, Eq.\ (\ref{prob_fcn}), evaluated for the two cases. Consistently with Ampere's law and the remarks above, more (less) vorticity must be crowded toward the origin for the higher (lower) energies. We should bear in mind that associated with each individual line vortex, there is an infinite positive self-energy. This is not included in what we are calling the ``mean-field energy,'' which is a sum of potential energies between pairs only. Nevertheless, choosing mean-field energies above that of the uniform distribution greatly enchances the ability of a given number of line vortices to strengthen the high-velocity tails: crowding the vortices together produces more opportunities for nearest neighbor events in the regions of enhanced mean-field vorticity.  Also, where there is a high probability density, we may expect a large value of the average velocity that is not attributable to nearest neighbor events.

Fig.\ \ref{fig5} displays the vorticity probability distribution at $r = 40.7$ for the ${\cal E} = 4{\cal E}_0$ case; this is inside the region of high radial probability density. Note the very large value of $\overline{w}$ and the associated large values of $w^2$. The probability of finding a vortex near this point is so high, in fact, that every single trial contained at least one nearest-neighbor event, so the corresponding graph with nearest neighbor events deleted has no data points in it, according to our previously-chosen criterion. We also observe that the nearest-neighbor formula (broken line) and the Gaussian (solid line) are not far apart for this case.

Figs.\ \ref{fig6}a,b are also for the high-energy case, but sample the velocity field at $r=114$, an intermediate value. Here we observe, as in the uniform vorticity density case, a noticeable high-velocity tail attributable to the nearest-neighbor events which disappears when those events are deleted. The much lower value of $\overline{w}=3.3$ is close to what was seen in the uniform vorticity case, and far lower than in Figs.\ \ref{fig5}a,b.  Not only the mean-fields, but the statistics of the fluctuations, are now strongly position-dependent.  This point is made even more strongly by looking at the velocity distribution at $r=399$, near the wall (Figs.\ \ref{fig7}a,b).  Here, where the probability distribution is very low, there is little velocity fluctuation ($\overline{w}=0.35$).  Here, the nearest-neighbor calculation is of severely limited applicability. The Gaussian is still present, as is the high-velocity tail, but the high-velocity tail does not disappear when the nearest neighbor events are deleted. The nearest neighbor formula derivation takes no account of the proximity of the wall, effectively assuming a rotational symmetry about the point of observation which is not even approximately fulfilled near the wall. The boundary condition begins to make itself strongly felt in this case, and it is not obvious how to include it in any theory. 

Turning now to the second set of trials, with ${\cal E}= {\cal E}_0 / 4 $, we consider the case where the probability is concentrated near the walls. We present the results of sampling at the radius $r=147$ (Figs.\ \ref{fig8}a,b). This is again an intermediate regime where the results are not greatly different from the uniform mean-vorticity case. Closer to the wall, the locally larger values of $p$ again diminish the differences between this case and the uniform $\omega$ case.

In summary, there are some strong qualitative similarities between the uniform and non-uniform mean field vorticity cases: the division into Gaussian plus high-velocity tail is usually applicable.  One principal quantitative difference is that the fluctuation level becomes more intense for the high-energy cases in those regions where the vorticity is concentrated. The mean velocity can also go up, and the mean field also becomes more intense. The overall fluctuation level goes up dramatically with mean-field energy. Though we do not have a theory for how fast it should go up, we can see from Fig.\ \ref{fig9} that it is considerably faster than linear. Fig.\ \ref{fig9} shows the mean field energy, normalized to the uniform mean-vorticity values, as a function of mean-field energy, for the three values of mean-field energy considered. Adding points to this graph is an expensive and time-consuming activity, but would seem to be a worthwhile undertaking.  The significantly noisier high-energy states for the system is something that will be characteristic of the ideal line vortex model but not for continuum models of a fluid.

\section{Closing Remarks}
\label{sec-conclusions}

We have investigated numerically the statistics of the Eulerian velocity field in two dimensional flows generated by a large number of ideal, parallel, line vortices inside an axisymmetric rigid boundary. This is a dynamical system the statistical mechanics of which have been interesting to investigate in their own right, and which also seem to have implications, not fully elucidated, for two-dimensional viscous continuum flows \cite{matthaeus,montmatt,schep}.  By considering the numerical effects of ``near neighbors'' and their contributions to the velocity fields at fixed spatial points, we have to a considerable degree confirmed the hypothesis of Kuvshinov and Schep \cite{kuvshinov} that the observed non-Gaussian, approximately third power ``tails'' in the velocity field distribution are due to these near neighbor events. These tails coexist with a ``bulk'' Gaussian distribution at lower velocities. 

The phenomenon of non-Gaussian high-velocity tails in measurement and computation of three-dimensional continuum fluid turbulence has been observed before (e.g., Vincent and Meneguzzi \cite{vincent91}; see also Jimenez \cite{jimenez}).  In computations, also simultaneously visible have been concentrated vortex configurations that have variously been called ``tubes,'' ``worms'' or ``spaghetti,'' since they are longer by a considerable amont in one dimension than they are in the other two. Accounting for these configurations has been an important problem. It is difficult not to imagine that the one might be responsible for the other.  That is, we suggest that the non-Gaussian tails are a signature of  physically proximate strong, tubular vortices which are enough like ``line" vortices that they account for the tails in three dimensions in the manner observed here in pure two-dimensional form.

A second part of the investigation has been motivated by the recognition that pairwise interaction energies, summed over all the pairs of an assembly of identical line vortices, provides a finite integral of the motion that can be set at any value, and determines as much about the thermal equilibria that are possible as energy usually does for conservative statistical-mechanical systems. The non-uniform mean-field distribution which results can impact the microscopic fluctuation distribution for a fixed number of vortices by creating more (and therefore noisier) regions where ``near neighbors'' reside. Such an effect will undoubtedly enhance transport properties, such as the coefficient of self diffusion \cite{taylor,mont72}, because of the larger random velocities which result. 

It would be of interest to follow up these investigations with dynamical computations, in which an assembly of line vortices was moved around by its self-consistent velocity field, with an eye toward measuring two-time statistical correlations of Eulerian velocity fields, diffusion and decay rates. Measured coefficients of self-diffusion may be determined numerically, and may be found to depend fundamentally on the mean-field energy and consequent temperature that characterize a vortex equilibrium and not to be representable by any ``universal'' formula. Much earlier computations and theories for ideal line vortex dynamics \cite{jimenez,lundgren,montjoyce73} showed unexpected late implications for Navier-Stokes fluid turbulence in two dimensions \cite{matthaeus,montmatt}. Standard ``homogeneous turbulence'' theories were shown to be very poor predictors for the late-time states of turbulent fluids in two dimensions, once this step was taken. We may speculate that the present considerations, which extend Holtsmark statistics beyond the spatially uniform case, might substantially revise, for example, the magnitudes of transport coefficients that are often assigned to such diverse systems as galaxies or globular clusters \cite{chandra} and dilute magnetized plasmas \cite{taylor,mont72}.

\acknowledgments
One of us (T.S.L.) was supported under a Waterhouse Research Grant from Dartmouth College. The other (D.C.M.) would like to express gratitude for hospitality in the Fluid Dynamics Laboratory at the Eindhoven Technical University in the Netherlands, where part of this work was carried out.


%
%
\begin{figure}
\centering
\includegraphics[angle=0, width=6.0in, height=6.0in]{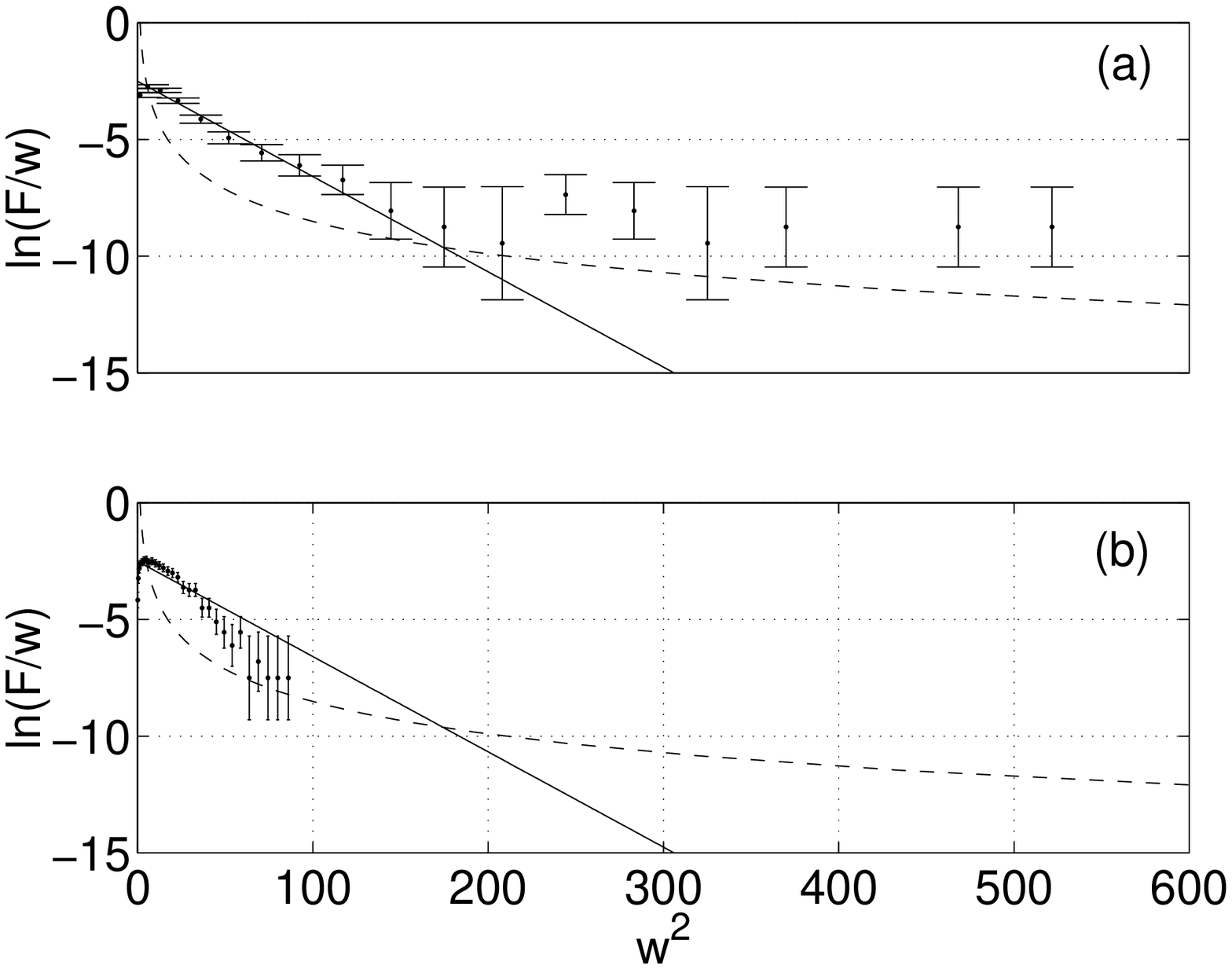}
\caption{Plot of $\ln (F/w)$ vs.\ $w^2$ at $r=0$ for the uniform case. The upper graph (a) contains nearest neighbor events. The lower graph (b) has nearest neighbor events deleted. The solid line represents a best fit Gaussian ($ \overline{w} = 3.5$). The dashed line is the analytical expression for the nearest neighbor effects.}
\label{fig1}
\end{figure}

\begin{figure}
\centering
\includegraphics[angle=0, width=6.0in, height=6.0in]{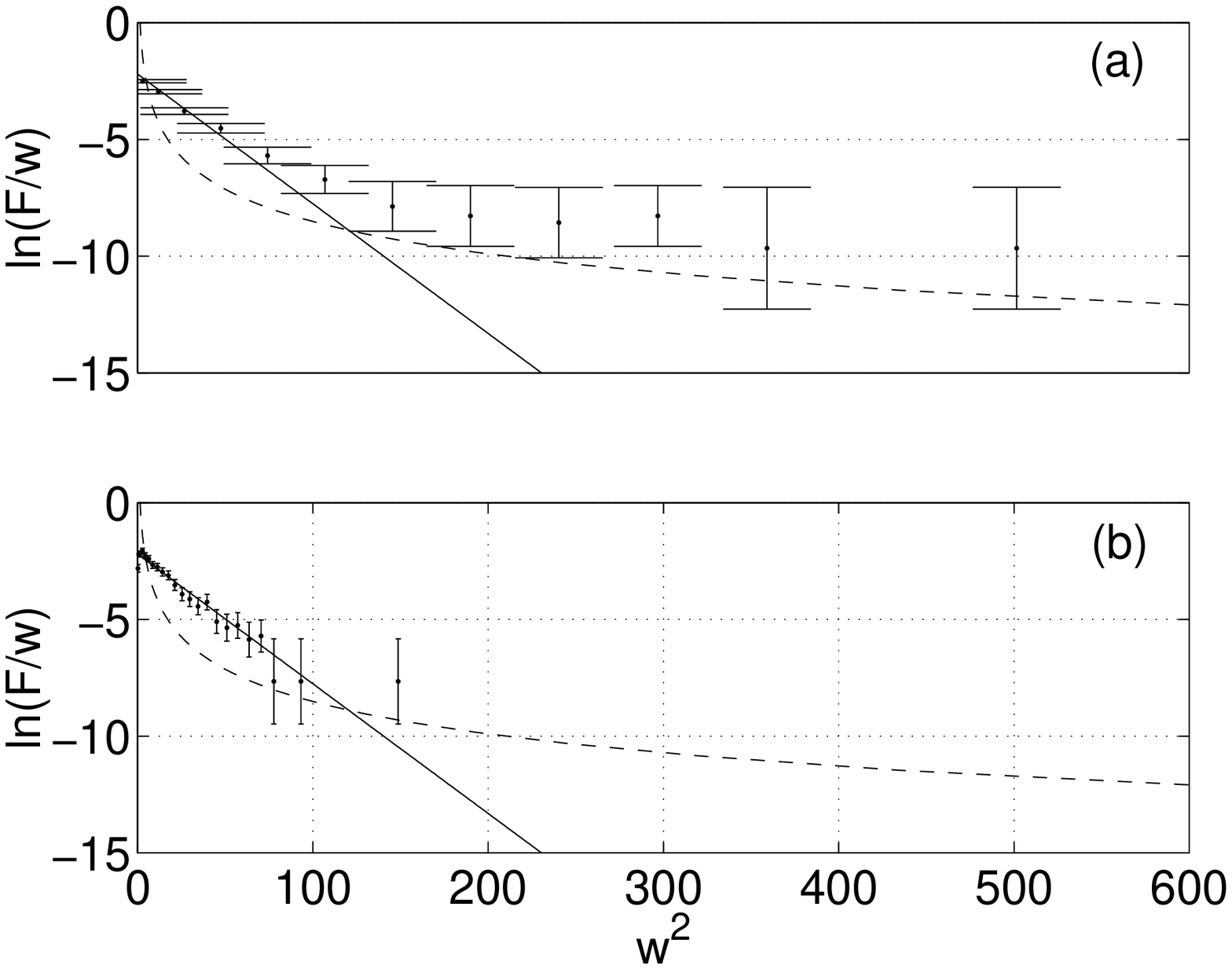}
\caption{Plot of $\ln (F/w)$ vs.\ $w^2$ at $r=399$ for the uniform case. The upper graph (a) contains nearest neighbor events. The lower graph (b) has nearest neighbor events deleted. The solid line represents a best fit Gaussian ($ \overline{w} = 3.0$). The dashed line is the analytical expression for the nearest neighbor effects.}
\label{fig2}
\end{figure}

\begin{figure}
\centering
\includegraphics[angle=0, width=6.0in, height=6.0in]{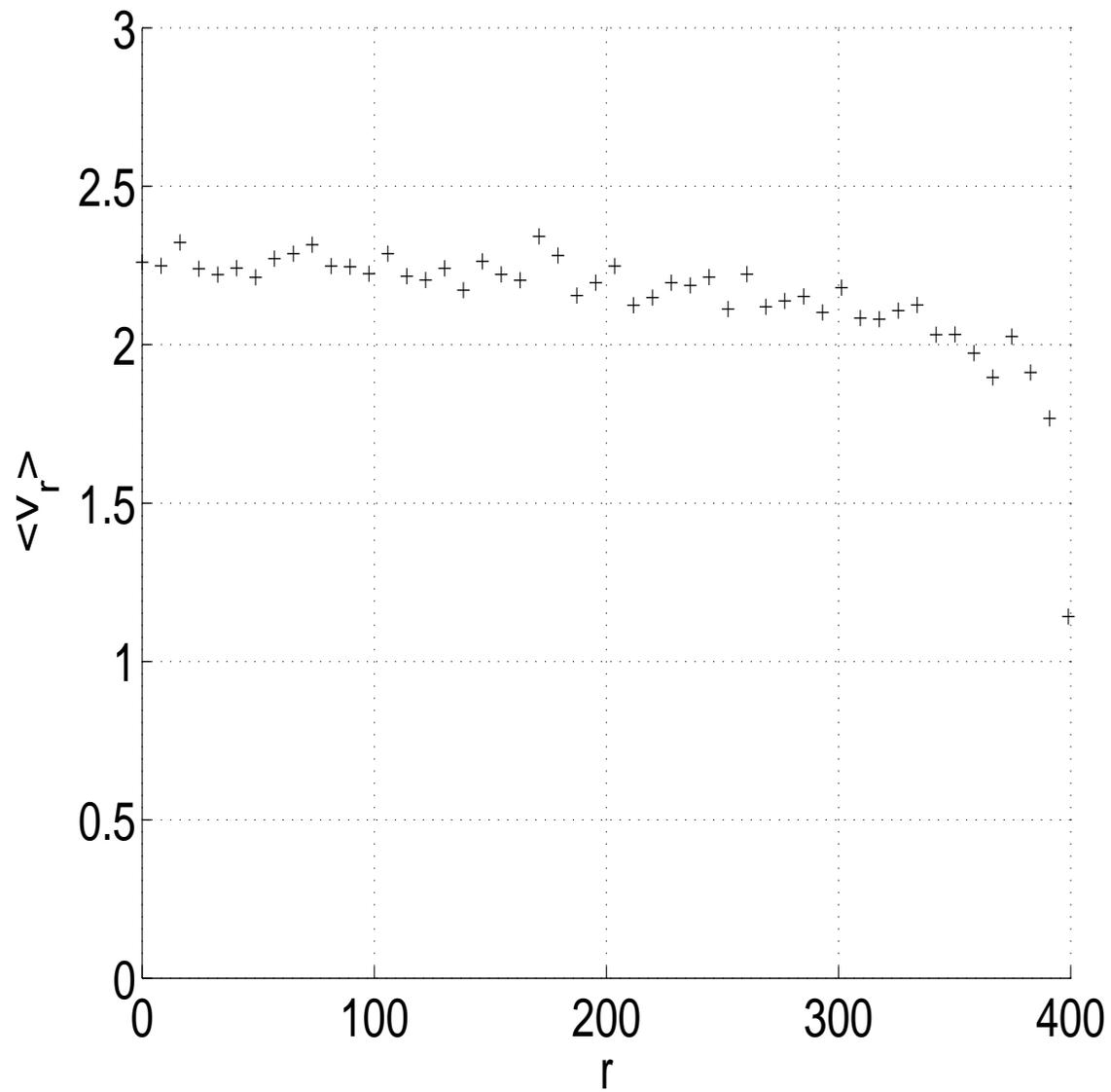}

\caption{Plot of $<v_r>$ vs.\ $r$. Notice the sharp drop towards zero near the wall at $r=R=400 $. This is evidence of a relatively thin boundary layer near the wall.}
\label{fig3}
\end{figure}

\begin{figure}
\centering
\includegraphics[angle=0, width=6.0in, height=6.0in]{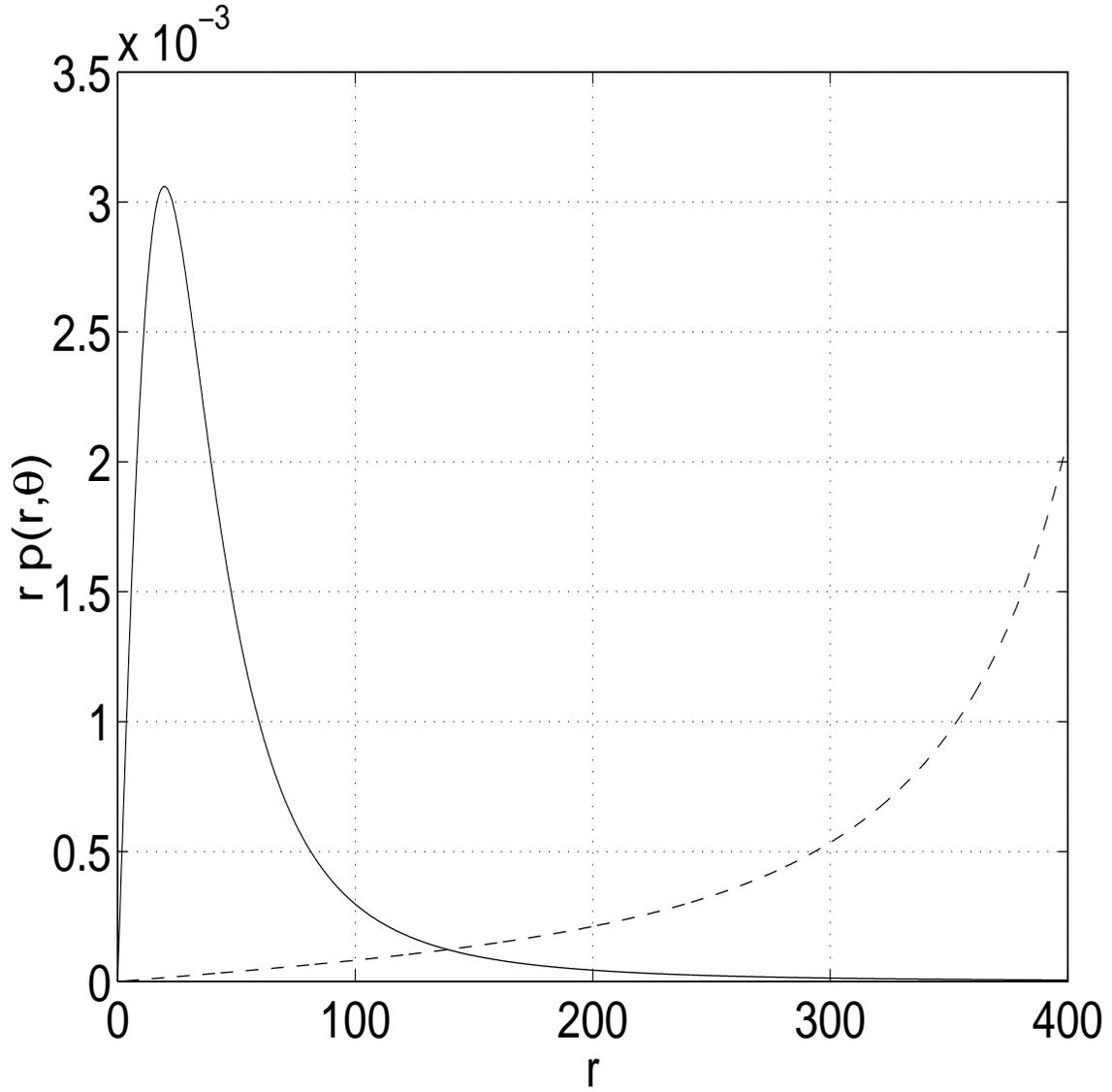}
\caption{Plot showing $r p(r, \theta )$ vs. $r$. The solid line is the case where ${\cal E} = 4{\cal E}_0$. The dashed line is the case where ${\cal E} = {\cal E}_0 / 4$. }
\label{fig4}
\end{figure}

\begin{figure}
\centering
\includegraphics[angle=0, width=6.0in, height=6.0in]{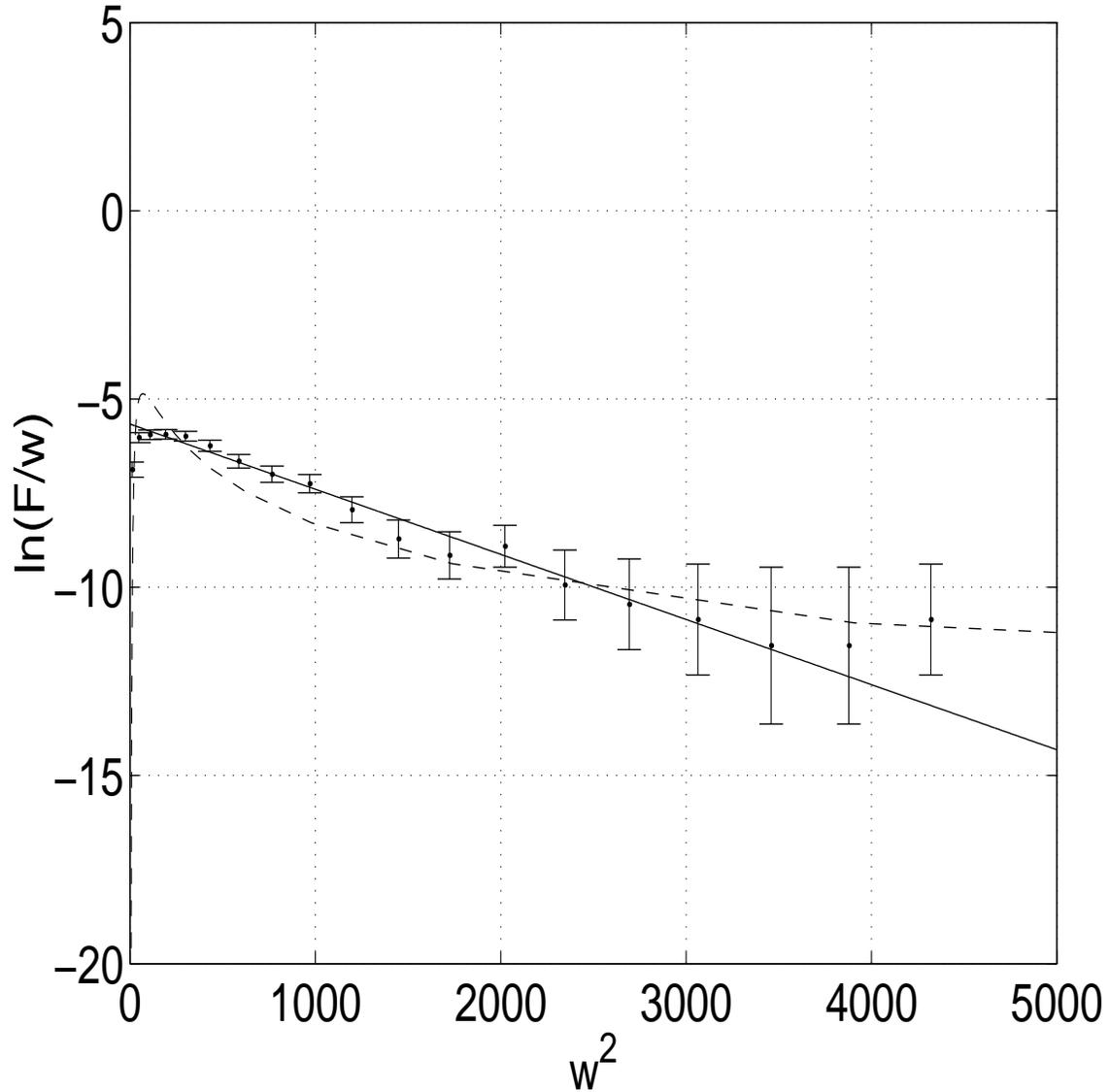}
\caption{Plot of $\ln (F/w)$ vs.\ $w^2$ at $r=40.7$ for the ${\cal E} = 4{\cal E}_0$ case. Here, every point has a nearest neighbor event recorded and thus, the corresponding graph with nearest neighbor events deleted contains no points. The solid line represents a best fit Gaussian ($ \overline{w} = 17$). The dashed line is the analytical expression for the nearest neighbor effects.}
\label{fig5}
\end{figure}

\begin{figure}
\centering
\includegraphics[angle=0, width=6.0in, height=6.0in]{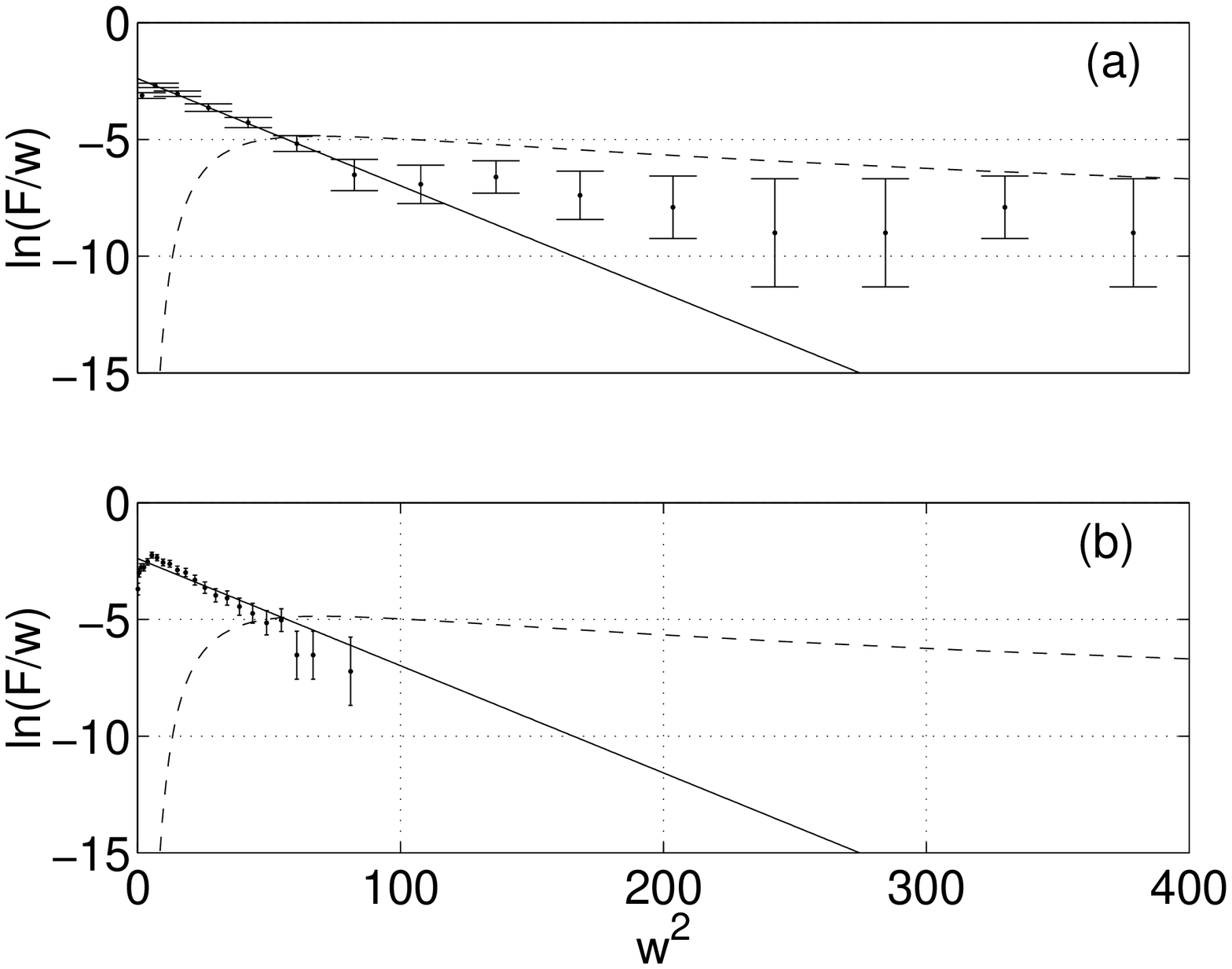}
\caption{Plot of $\ln (F/w)$ vs.\ $w^2$ at $r=114$ for the ${\cal E} = 4{\cal E}_0$ case. The upper graph (a) contains nearest neighbor events. The lower graph (b) has nearest neighbor events deleted. The solid line represents a best fit Gaussian ($ \overline{w} = 3.3$). The dashed line is the analytical expression for the nearest neighbor effects.}
\label{fig6}
\end{figure}

\begin{figure}
\centering
\includegraphics[angle=0, width=6.0in, height=6.0in]{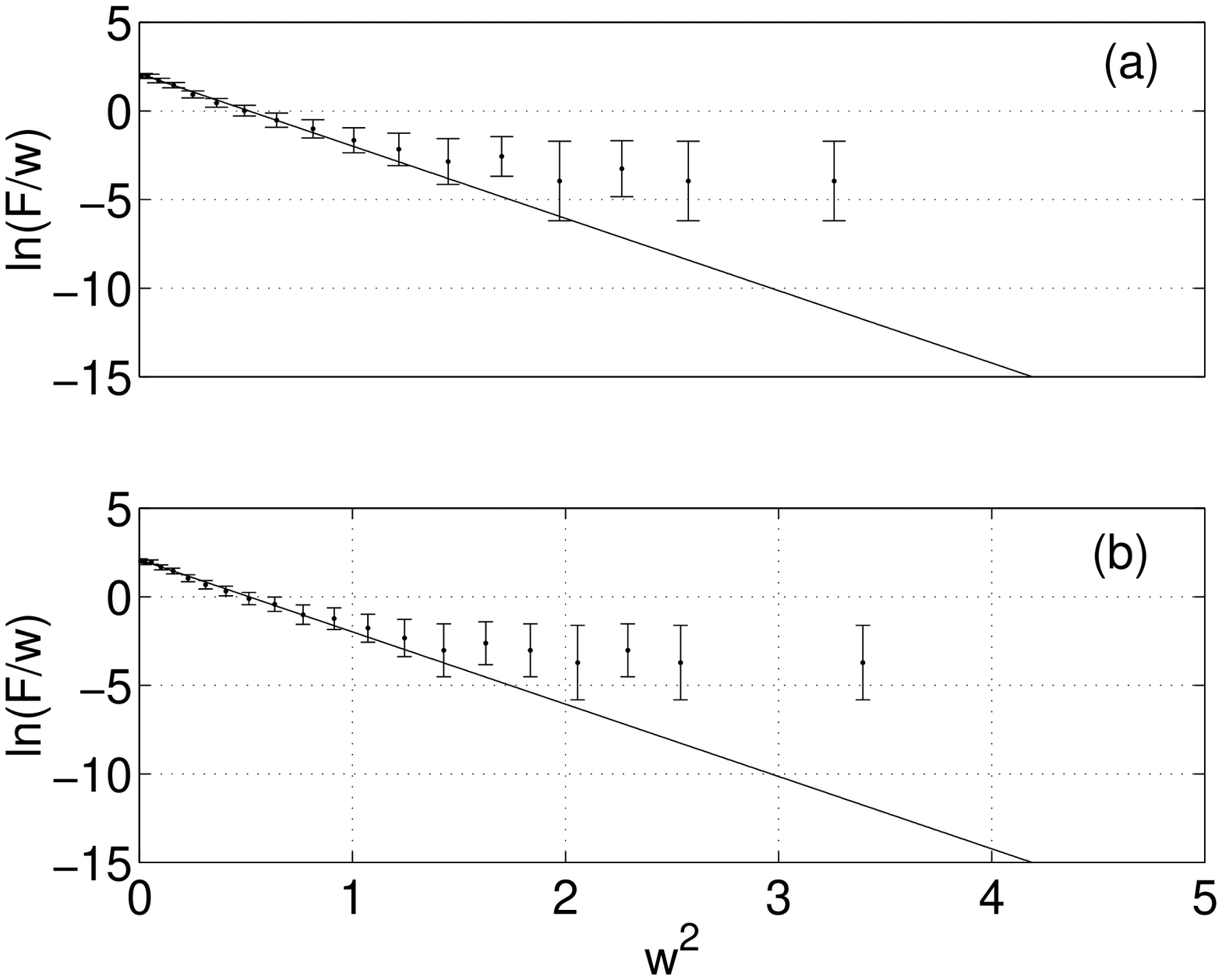}
\caption{Plot of $\ln (F/w)$ vs.\ $w^2$ at $r=399$ for the ${\cal E} = 4{\cal E}_0$ case. The upper graph (a) contains nearest neighbor events. The lower graph (b) has nearest neighbor events deleted. The solid line represents a best fit Gaussian ($ \overline{w} = 0.35$). The dashed line is the analytical expression for the nearest neighbor effects.}
\label{fig7}
\end{figure} 

\begin{figure}
\centering
\includegraphics[angle=0, width=6.0in, height=6.0in]{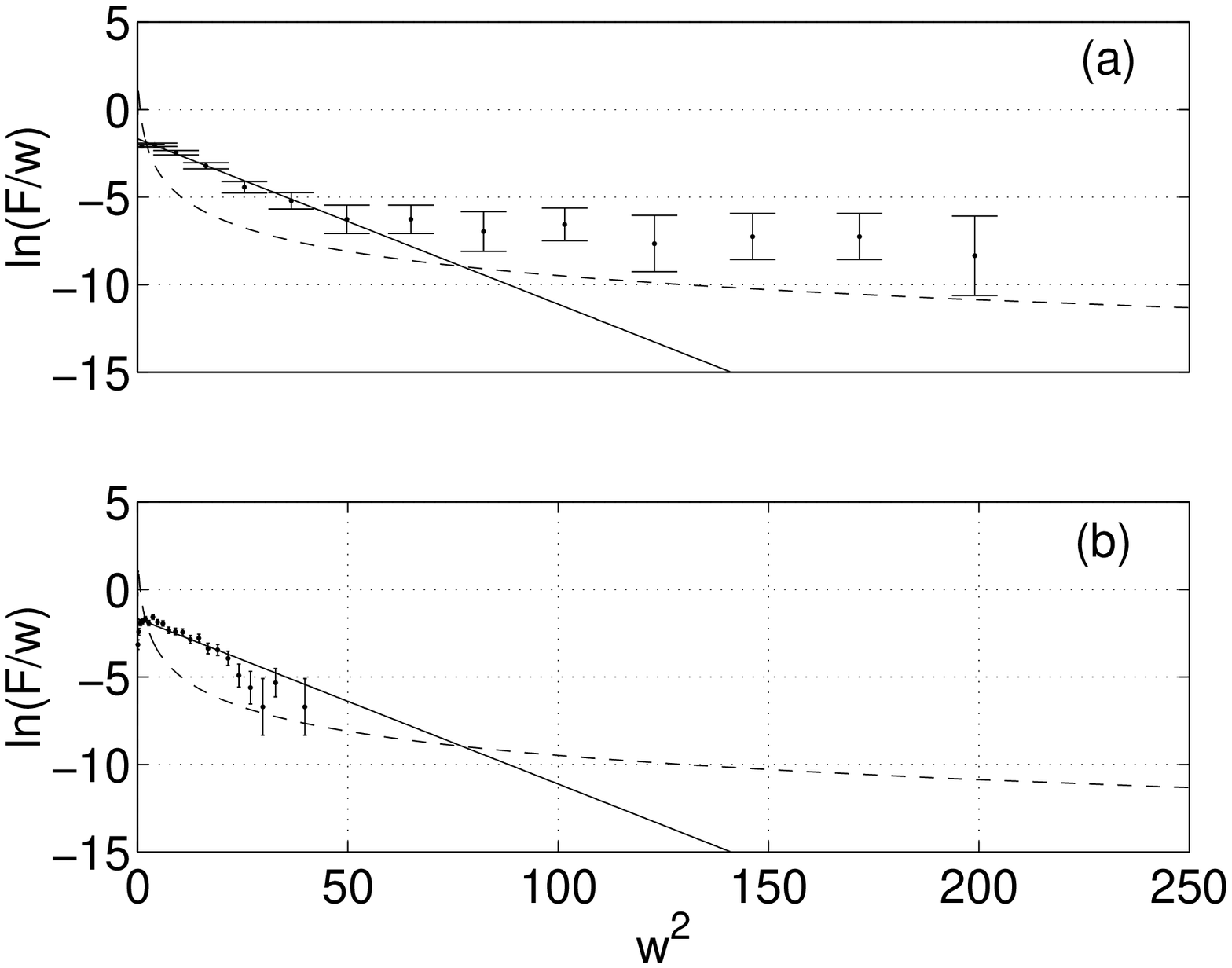}
\caption{Plot of $\ln (F/w)$ vs.\ $w^2$ at $r=147$ for the ${\cal E} = {\cal E}_0 / 4$ case. The upper graph (a) contains nearest neighbor events. The lower graph (b) has nearest neighbor events deleted. The solid line represents a best fit Gaussian ($ \overline{w} = 2.3$). The dashed line is the analytical expression for the nearest neighbor effects.}
\label{fig8}
\end{figure}

\begin{figure}
\centering
\includegraphics[angle=0, width=6.0in, height=6.0in]{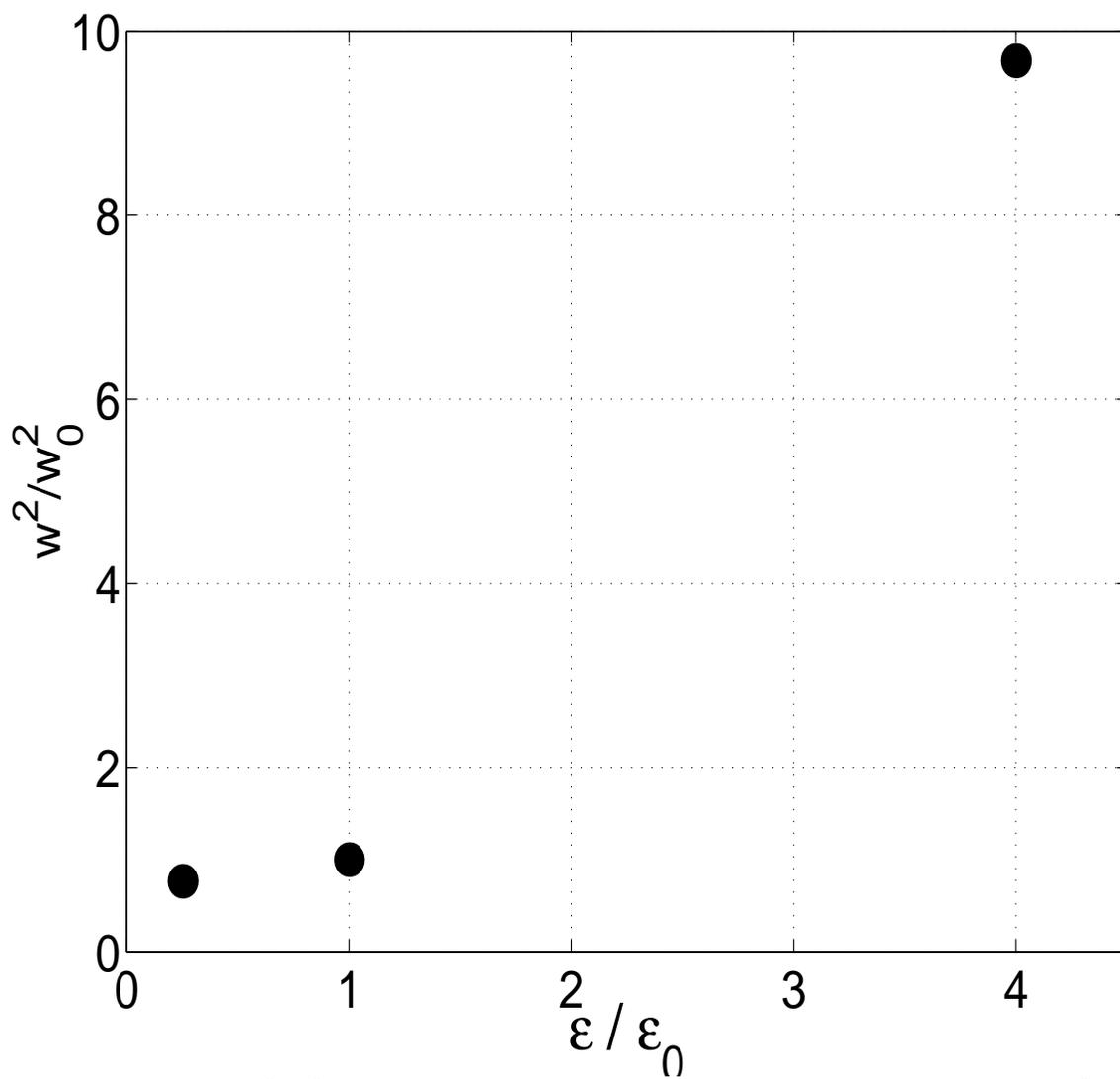}
\caption{Average of $w^2 / w^{2}_0$ for every point sampled plotted as a function of ${\cal E} / {\cal E}_0$ where $w^{2}_0=23.3$ is the value at ${\cal E}_0$.}
\label{fig9}
\end{figure}

%
%

\end{document}